\documentclass[preprint, aps]{revtex4}

\usepackage{graphicx}

\begin{document}




\title{Vector Field and Inflation}

\author{Seoktae Koh}
\address{Institute of Theoretical Physics, Chinese Academy of Sciences, \\
P.O. Box 2735, Beijing, 100190, China \\
skoh@itp.ac.cn}



\begin{abstract}
We have investigated if the vector field can give rise to an accelerating 
phase in the early universe. 
We consider a timelike vector 
field with a general quadratic kinetic term in order to preserve
an isotropic background spacetime. The vector field potential
is  required to satisfy the three minimal conditions
for successful inflation: 
i) $\rho>0$, ii) $\rho+3P < 0$ and iii) the slow-roll conditions. 
As an example,
we consider the massive vector potential and small field type potential 
as like in scalar driven inflation. 

\end{abstract}

\maketitle

\section{Introduction} 
Vector field inflation was firstly realized  in Ref.~\cite{Ford:1989me},
however it had a fine tuning problem 
for a successful inflationary expansion.
 Recently
a successful vector inflation model is constructed \cite{Golovnev:2008cf}
using an orthogonal triple vector set and nonminimal coupling
to gravity. In addition, if a vector field exists, 
anisotropic inflation can be possible 
\cite{Kanno:2008gn} against the cosmic no hair theorem.
The vector field is also considered to explain the 
dark energy problem\cite{ArmendarizPicon:2004pm}. 
However, most of the vector field models are plagued by instabilities
or by the  ghost modes
\cite{Himmetoglu:2008zp,Carroll:2008em}. 

In this work, we consider a timelike vector field instead of
a spacelike vector field\cite{Golovnev:2008cf,Kanno:2008gn} 
in order to preserve an isotropic background spacetime
and investigate if this vector field can give rise to an 
accelerating phase. Unlike spacelike vector field
inflation\cite{Golovnev:2008cf}, the nonminimal coupling to gravity with a
timelike vector field
could not help to generate an inflationary period\cite{Koh:2009vm}.
Instead, we try to find a vector field potential which can
give rise to a successful inflationary expansion.

This paper is organized as follows: in Section \ref{sec:background}
we describe our model and derive the background equations of motion.
We discuss about the conditions for the vector field potential 
to be satisfied
in Section \ref{sec:potential} and apply to the several potentials.
We conclude in Section \ref{sec:con}.

\section{Models and the equations of motion}\label{sec:background}
We begin with an action of a vector field which has
a general quadratic kinetic term\cite{Jacobson:2000xp}
\begin{eqnarray}
\mathcal{S} = \int d^4 x \sqrt{-g}
\left[\frac{1}{16\pi G}R -\frac{1}{2}
{K^{\mu\nu}}_{\rho\sigma}\nabla_{\mu}A^{\rho}\nabla_{\nu}A^{\sigma}
- V(\xi)
\right],
\label{oriaction}
\end{eqnarray}
where
\begin{equation}
{K^{\mu\nu}}_{\rho\sigma} = \beta_1 g^{\mu\nu}g_{\rho\sigma}
+\beta_2 \delta^{\mu}_{\rho} \delta^{\nu}_{\sigma}
+\beta_3 \delta^{\mu}_{\sigma} \delta^{\nu}_{\rho},
\end{equation}
$\xi = A_{\mu}A^{\mu} = A^2$,
and $V$ is a vector field potential. For a massive vector field,
$V(\xi) = \frac{1}{2}m^2 \xi$. We can consider the general form of 
the potential in addition to the linear massive potential 
\cite{Ford:1989me,koh09}.

Here $\beta_i$'s are dimensionless parameters. 
The standard Maxwell kinetic term, $-\frac{1}{4}F_{\mu\nu}F^{\mu\nu}$ where
$F_{\mu\nu} = \partial_{\mu}A_{\nu} - \partial_{\nu}A_{\mu}$, corresponds to
$\beta_1 = -\beta_3 = 1$, and  $\beta_2 =0$.
In Ref.~\cite{Carroll:2008em}, they have categorized into several 
models in terms of $\beta_i$'s. Among those models, 
 $\beta_1=\beta_T$ model, $\beta_T =0$ model and $\beta_1 = 0$ model, 
where $\beta_T = 
\beta_1+\beta_2+\beta_3$, seem to be free from linear instabilities
or negative-energy ghosts.

In our previous work\cite{Koh:2009vm},
we have considered  $\beta_1=\beta_2 = -\beta_3 = 1$ case
  and the kinetic Lagrangian density of the
vector field becomes
\begin{eqnarray} 
\label{lagrangian1}
\mathcal{L}_{A} = -\frac{1}{4}F_{\mu\nu}F^{\mu\nu} - \frac{1}{2}
(\nabla_{\mu}A^{\mu})^2.
\end{eqnarray}
This kinetic model is dubbed as a ``sigma model'' 
in Ref.~\cite{Carroll:2008em,Carroll:2009en}
and is stable under the small perturbations.
\footnote{This kinetic term is equivalent to
$\mathcal{L}_A = -\frac{1}{2}\nabla_{\mu}A_{\nu}\nabla^{\mu}A^{\nu}
-\frac{1}{2}R_{\mu\nu}A^{\mu}A^{\nu}$ up to the total derivative terms where
we have used $[\nabla_{\mu}, \nabla_{\nu}]A^{\mu} = R_{\mu\nu}A^{\mu}$.}

Another interesting case is $\beta_1 = \beta_3 = 0, ~\beta_2 =1$ or
$\beta_T =\beta_2 = 1$ and the model is defined by the form
\begin{eqnarray}
\mathcal{L}_A = -\frac{1}{2} (\nabla_{\mu} A^{\mu})^2.
\label{lagrangian2}
\end{eqnarray}
The background dynamics of a vector field
with this Lagrangian density  can not be discriminated from that
of (\ref{lagrangian1}), which will be shown later.

We can also think of $\beta_1 = -\beta_2 = \beta_3 = 1$ 
 case which is given by the form 
\begin{eqnarray}
\mathcal{L}_{A} = -\frac{1}{2}\nabla_{\mu}A_{\nu} \nabla^{\mu}
A^{\nu} + \frac{1}{2} R_{\mu\nu}A^{\mu}A^{\nu}.
\label{lagrangian3}
\end{eqnarray}
Unlike (\ref{lagrangian1}) and (\ref{lagrangian2}), this model
has $\beta_T = -\beta_2$. This model may have a
superluminal mode\cite{Carroll:2009en}, 
which means that the dynamical degree of freedom 
can propagate faster than light and lead to instabilities.

The equation of motion obtained by varying the action (\ref{oriaction})
with respect to $A_{\mu}$ is 
\begin{equation} \label{bgeom}
\nabla_{\mu}{J^{\mu}}_{\nu} = 2\frac{dV}{d\xi}A_{\nu},
\end{equation}
where
\begin{equation}
{J^{\mu}}_{\nu} = {K^{\mu\rho}}_{\nu\sigma}\nabla_{\rho}A^{\sigma}
=\beta_1 \nabla^{\mu}A_{\nu} +\beta_2 \delta^{\mu}_{\nu}\nabla_\rho A^{\rho}
+\beta_3 \nabla_{\nu}A^{\mu}.
\end{equation}

And by varying the action with respect to $g_{\mu\nu}$, 
one obtains Einstein equations
\begin{eqnarray}
R_{\mu\nu} &-& \frac{1}{2}g_{\mu\nu} R 
= 8\pi G T_{\mu\nu}, \\ 
T_{\mu\nu} &=& 
\beta_1(\nabla_{\mu}A^{\rho}\nabla_{\nu} A_{\rho} -\nabla^{\rho} A_{\mu}
\nabla_{\rho} A_{\nu})
+ \frac{1}{2} [ \nabla_{\lambda}(A_{\nu} {J^{\lambda}}_{\mu} 
 + A_{\mu}{J^{\lambda}}_{\nu})  \nonumber \\
& & +\nabla_{\lambda}(A^{\lambda}J_{\mu\nu}+A^{\lambda}J_{\nu\mu})
-\nabla_{\lambda}(A_{\nu}{J_{\mu}}^{\lambda}+A_{\mu}{J_{\nu}}^{\lambda}) ] 
\nonumber \\
& & -\frac{1}{2}g_{\mu\nu} {K^{\alpha\beta}}_{\rho\sigma}\nabla_{\alpha}A^{\rho}
\nabla_{\beta}A^{\sigma}
+ 2\frac{dV}{d\xi}A_{\mu}A_{ \nu}  -g_{\mu\nu}V(\xi). 
\end{eqnarray}

\begin{figure}[tb]
\includegraphics[width=0.6\textwidth]{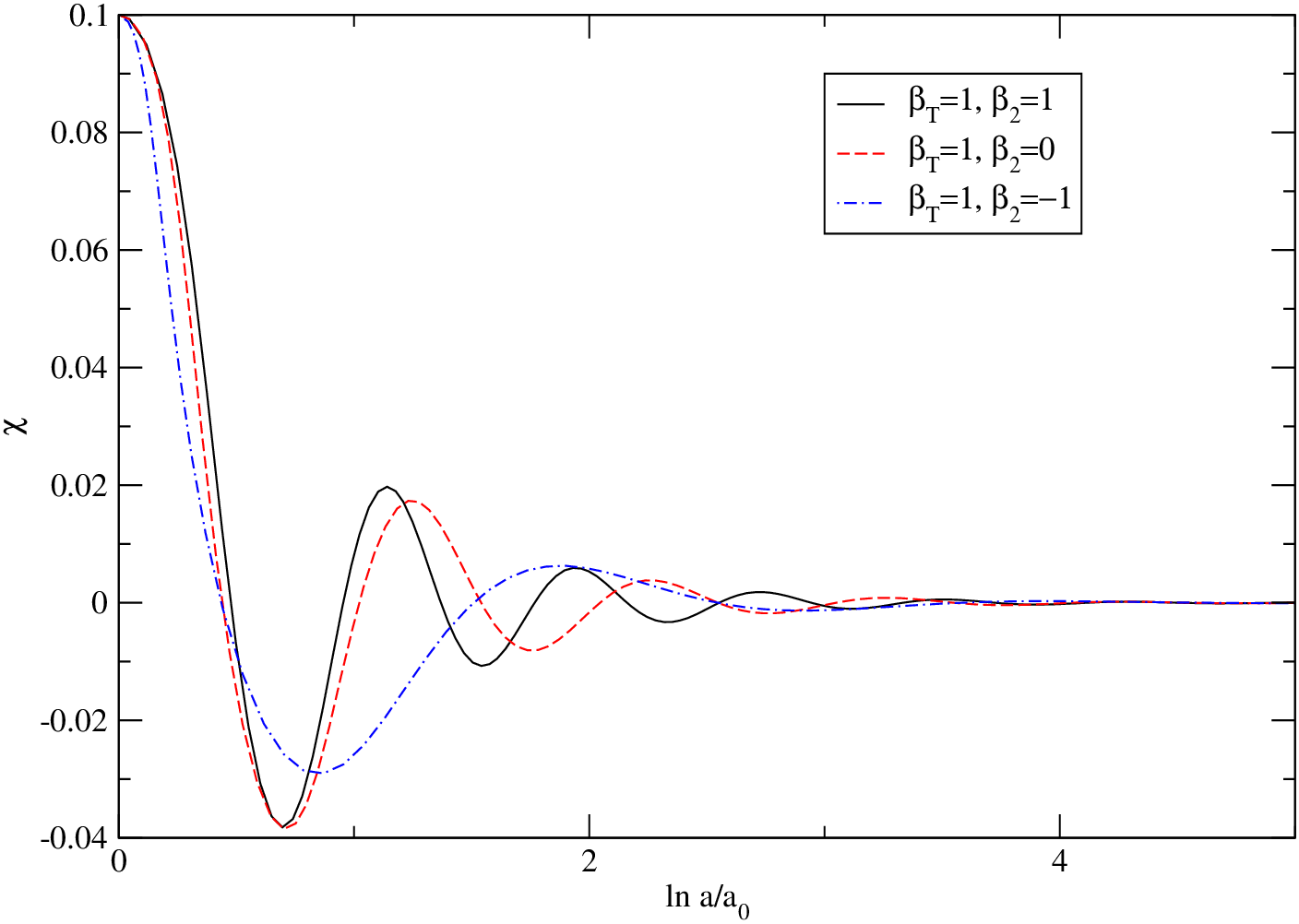}
\caption{The evolutions of $\chi$ are plotted 
with different $\beta_T$ and $\beta_2$. We set to
$\chi_i = 0.1 m_{pl}$}
\label{fig1}
\end{figure}

In order to preserve an isotropic background spacetime, we will use
a timelike vector field ($\xi \equiv A_{\mu}A^{\mu} < 0$) and then take as
$A_{\mu}^{(0)}(t) = (A_0(t),\vec{0}) \equiv (\chi(t),\vec{0})$.
In the spatially flat FRW metric
\begin{equation}
ds^2 = -dt^2 + a^2(t) \gamma_{ij}dx^i dx^j,
\label{bgmetric}
\end{equation}
the field equation of $\chi$ and Einstein equations 
can be expressed as
\begin{eqnarray}
\ddot{\chi} &+& 3H\dot{\chi} + \left[
3\beta_2 \dot{H} + 3(\beta_2-\beta_T)H^2 
+ 2\frac{dV}{d\xi} \right]\frac{1}{\beta_T}\chi = 0, 
\label{feom} \\
\left(\frac{\dot{a}}{a}\right)^2 &=& 
\frac{8\pi G}{3} \Biggl[ 
-\beta_T(\chi \ddot{\chi} +\frac{1}{2}\dot{\chi}^2 +3H\chi \dot{\chi}
-\frac{3}{2}H^2 \chi^2)  \nonumber \\
& &~~~~ -3\beta_2 (H\chi\dot{\chi} +\dot{H} \chi^2 + 2H^2 \chi^2)
+ 2\frac{dV}{d\xi}\chi^2 + V \Biggr],
\label{friedman}\\
\dot{H} &=& 4\pi G \Biggl[ (\beta_T-\beta_2)(
\chi\ddot{\chi} +\dot{\chi}^2+H\chi\dot{\chi} -\dot{H}\chi^2
-3H^2 \chi^2 )
-2\frac{dV}{d\xi}\chi^2 \Biggr] .
\label{einij}
\end{eqnarray}
Here we have assumed $\beta_{T} \equiv \beta_1+\beta_2 +\beta_3 \neq 0$
for the timelike vector field to 
behave as a dynamical degree of freedom. The background evolutions of
the vector field in (\ref{feom}),
(\ref{friedman}), and (\ref{einij}) 
are determined  only by $\beta_T$ and $\beta_2$.
This means that the models with the Lagrangian density (\ref{lagrangian1})
and (\ref{lagrangian2}) can not be distinguished from the background dynamics.
In order to distinguish between these two models, we need to consider
the linear perturbations.
We plot the evolutions of the vector field with $\beta_T$ and
$\beta_2$ for $V(\xi) = \frac{1}{2}m^2 \xi$  in Fig. \ref{fig1}.

\section{Can inflation be possible?}\label{sec:potential}

\begin{figure}[tb]
\includegraphics[width=0.65\textwidth]{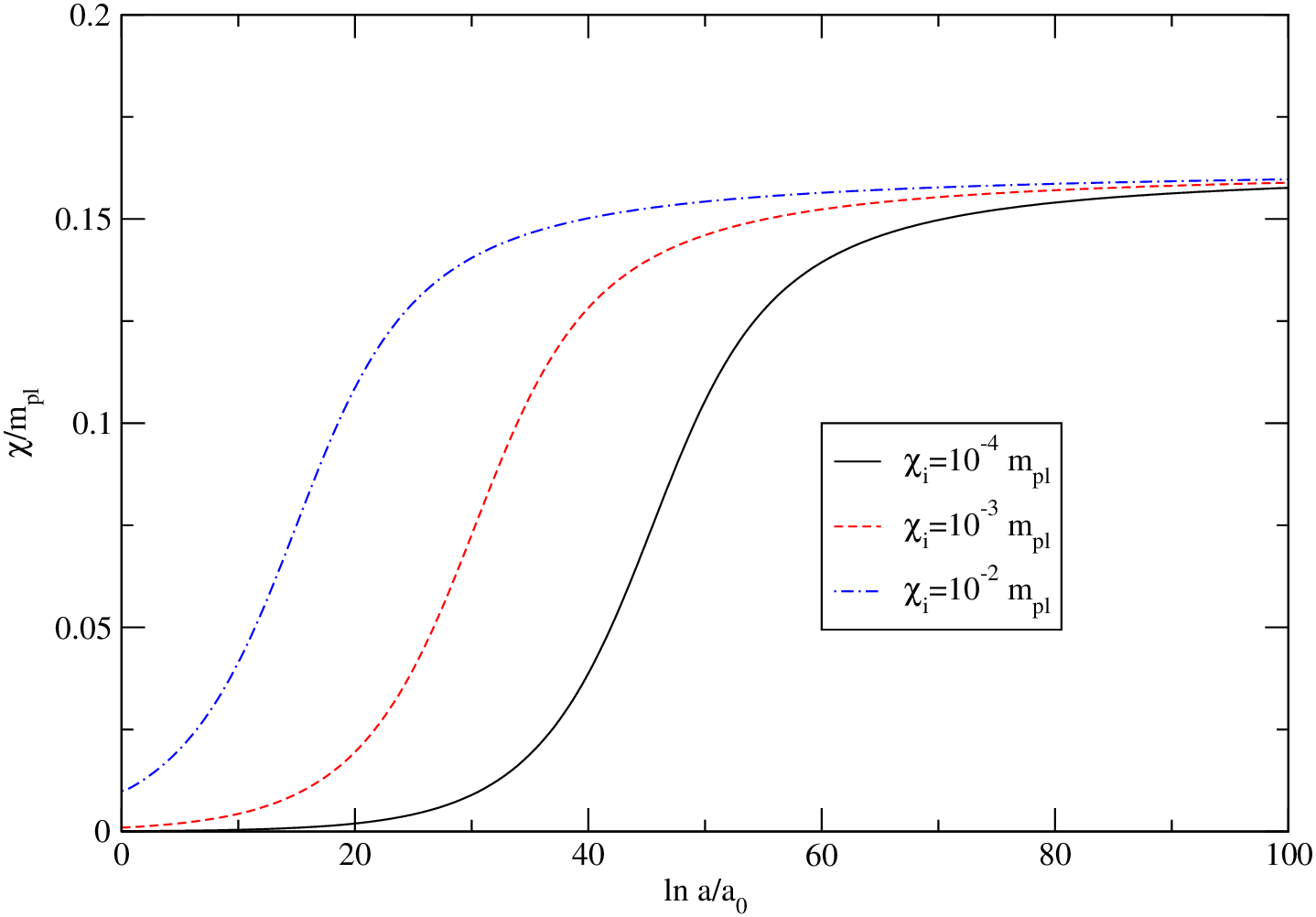}
\caption{The evolutions of vector field are depicted
depending on $\chi_i$
 for $V = V_0 - \frac{1}{2}m^2 \xi$. We set to $\mu =m_{pl}$.}
\label{fig2}
\end{figure}

In this section, we check whether the timelike vector field 
can give rise to the period of inflationary expansion.
In order to generate successful slow-roll inflation, slow-roll conditions
 ($\ddot{\chi} \ll  3H\dot{\chi}$,
$\dot{\chi}^2/2  \ll V(\xi)$ and $\dot{H} \ll H^2$) should be satisfied.
From (\ref{feom}), we can define the effective mass of the vector field as
\begin{eqnarray}
m_{eff} \equiv \frac{1}{\beta_T}\left(3\beta_2 \dot{H} + 3(\beta_2 -\beta_T)H^2
+2\frac{dV}{d\xi}\right).
\end{eqnarray}
If we assume that the slow-roll conditions are satisfied, then
$m_{eff} \simeq (3(\beta_2 - \beta_T)H^2 + 2dV/d\xi)/\beta_T$.
In order for the vector field to roll slowly over the effective potential,
$m_{eff}$ should be much smaller than $H^2$. But as long as $\beta_T 
\neq \beta_2$, $m_{eff} \sim H^2$ if $H^2 \gg dV/d\xi$ or
$m_{eff} \gg H^2$ if $H^2 \ll dV/d\xi$. 
Only for $\beta_T = \beta_2$, there would be a possibility
of $m_{eff} \ll H^2$ when $dV/d\xi \ll H^2$ is
satisfied.
Unless otherwise stated,
 we only focus on $\beta_T = \beta_2 = 1$ case in what follows.

In order to check the possibility of the inflationary expansion
 with the vector field,
we require the following three minimal conditions
one would expect a successful inflation model to 
satisfy: i)$H^2 \propto \rho > 0$
ii) $\ddot{a}/a \propto \rho + 3P < 0$ and iii) $|\dot{H}/H^2| \ll 1$.
First condition is required in order to have a positive energy density, 
second is for an accelerating
expansion and the final one is for slow-roll inflation.
Then we could find a vector field potential 
to satisfy these conditions.

We can summarize the above three requirements as
$\chi \ll m_{pl}, ~V +6 (dV/d\xi)\chi^2 > 0$ 
and $(|dV/d\xi|)\chi^2 \ll |V|$.\footnote{See  
details in Ref.~\cite{skoh09}.}
 With these conditions,  $H^2 \sim V$
and  the number
of efolds $N_e$ can be calculated as 
\begin{eqnarray}
N_{e} = \int H dt = \int \frac{H}{\dot{\chi}} d\chi
\sim \int \frac{4\pi V}{m_{pl}\chi|dV/d\xi|} d\chi.
\end{eqnarray}

We first consider the massive vector potential, 
$V = \frac{1}{2}m^2 \xi = -\frac{1}{2}m^2 \chi^2$, and apply 
the  above conditions to this potential.
Since $H^2 \sim \dot{H}$ and $m_{eff}^2 \sim H^2$
as shown in  Ref.~\cite{Koh:2009vm}, this potential can not satisfy 
the slow-roll conditions and thus
we can not obtain an accelerating  expansion. We plot 
the evolution of the vector field with this massive potential
in Fig. \ref{fig1}.

\begin{figure}[tb]
\includegraphics[width=0.65\textwidth]{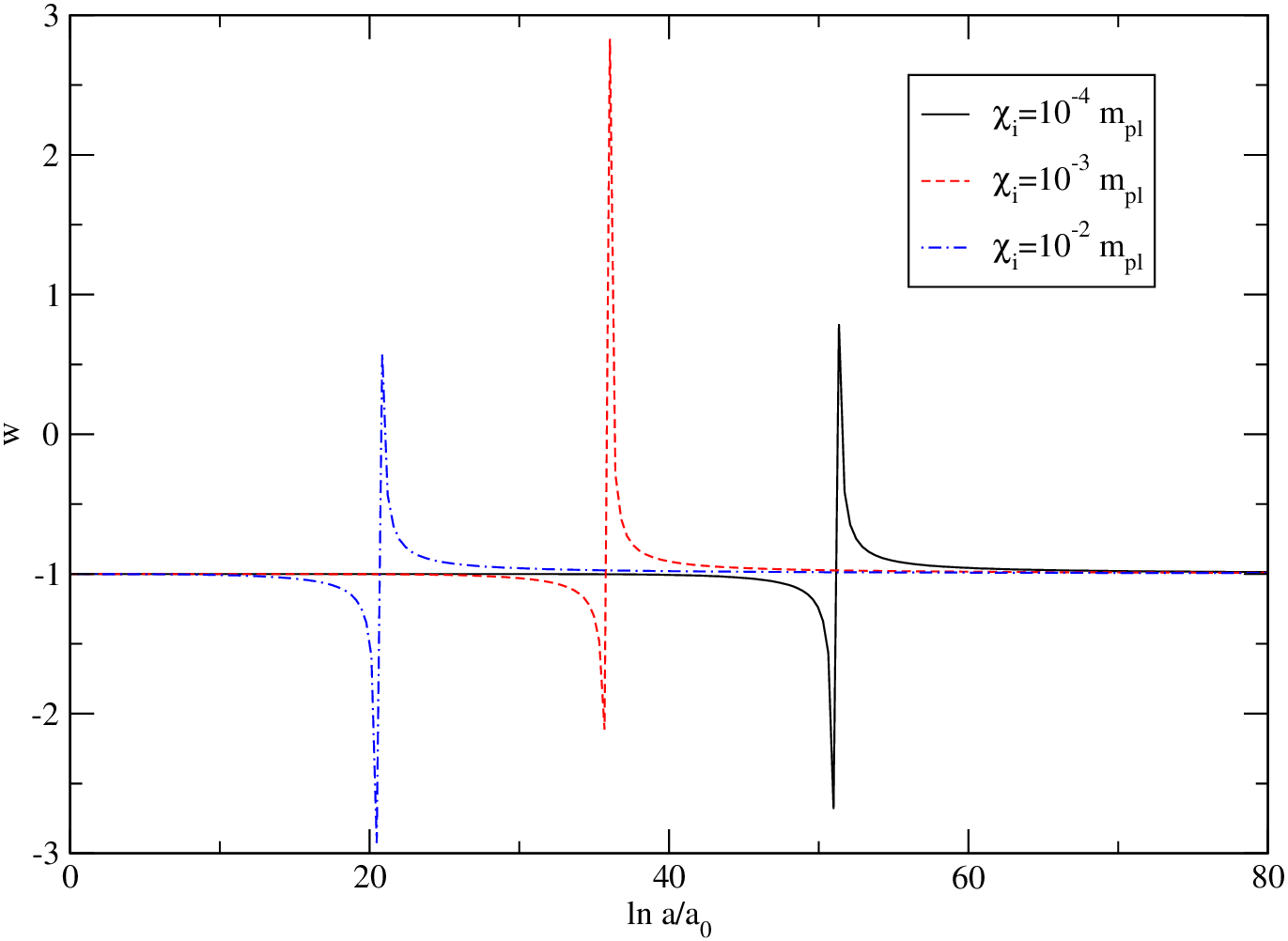}
\caption{The evolutions of equation of state parameter, $w$, are depicted
depending on $\chi_i$
 for $V = V_0 - \frac{1}{2}m^2 \xi$.}
\label{fig3}
\end{figure}

Next we consider the small field type potential 
as like in a scalar field inflation model
\begin{eqnarray}
V = \lambda (\xi - \mu^2)^2 \simeq V_0 - \frac{1}{2}m^2 \xi,
\end{eqnarray}
where $V_0 = \lambda \mu^4$ and $m^2 = 4\lambda \mu^2$ and
we have assumed  $\chi \ll \mu$ in the last expression.
This potential can satisfy all conditions as long as $\chi \ll \mu$.
We plot the evolution of $\chi$ and equation of state parameter, $w = P/\rho$,
 with the initial conditions of $\chi$ 
for this potential in Figs. \ref{fig2} and 
\ref{fig3}.
$\chi_i$ can be constrained from $N_e$ which 
is required to be larger than $50$ in order to fit to the observations.
But although the accelerating phase can occur, we need another mechanism
to exit to the standard cosmology.

\section{Conclusion and discussion}\label{sec:con}
We have investigated the vector field dynamics 
in the early universe and checked if the vector field can
give rise to an inflationary expansion. In order to preserve
an isotropic background spacetime, the timelike vector field is considered.
The vector field potential
is required to be $\rho > 0,~\rho+3P < 0$ and $|\dot{H}|\ll H^2$
in order to produce an inflationary expansion.
The small field type potential as like in scalar driven inflation, $V \simeq
V_0 +\frac{1}{2}m^2 \xi$, can give rise to an accelerating 
expansion and get a sufficient inflationary period by adjusting
the initial condition of the field. However, there still seems
to exist a problem to exit to the standard cosmology.

It would be interesting to calculate the scalar and tensor power spectrum
and to compare with the observations. It is also needed to check 
the instability in sub-Hubble length scale with our model.

\section*{Acknowledgments}
The author would like to thank
L. Ford, Seokcheon Lee and Hu Bin for useful discussions.
The author wishes to thank 
 Kunsan National University and Center for Quantum
Spacetime (CQUeST) at Sogang University for their warm hospitality
and financial support. 



\end{document}